\documentclass[aps,pra,showpacs,amssymb,nofootinbib,superscriptaddress,twocolumn,longbibliography]{revtex4-1}

\usepackage{epsfig,amsmath,amssymb,bm,epsf,graphicx,psfrag,bbm}
\usepackage{color, comment, verbatim}
\usepackage{ragged2e}
\usepackage{subfigure}
\usepackage{enumerate}
\usepackage{chngcntr}
\usepackage[T1]{fontenc}
\usepackage[english]{babel}
\usepackage[utf8]{inputenc}
\usepackage[colorlinks,breaklinks]{hyperref}
\usepackage{dsfont}
\usepackage{float}
\usepackage{tabularx}
\usepackage{graphicx}
\usepackage{amsbsy}
\usepackage{amsthm}

\usepackage{dsfont} 

\usepackage{tikz}
\usepackage[toc,page]{appendix}
\usepackage{titlesec}
\usepackage{xcolor}

\makeatletter
\newcommand\org@hypertarget{}
\let\org@hypertarget\hypertarget
\renewcommand\hypertarget[2]{%
  \Hy@raisedlink{\org@hypertarget{#1}{}}#2%
  }
\makeatother
\hypersetup{
	bookmarksnumbered,
	pdfstartview={FitH},
	citecolor={darkgreen},
	linkcolor={darkred},
	urlcolor={darkblue},
	pdfpagemode={UseOutlines}}
\definecolor{darkgreen}{RGB}{50,190,50}
\definecolor{darkblue}{RGB}{0,0,190}
\definecolor{darkred}{RGB}{238,0,0}

\usepackage[framemethod=tikz]{mdframed}
\definecolor{mycolor}{rgb}{0.122, 0.435, 0.698}
\definecolor{mycolor2}{RGB}{112, 48, 160}
\newmdenv[innerlinewidth=0.5pt, roundcorner=4pt,linecolor=mycolor,innerleftmargin=6pt,
innerrightmargin=6pt,innertopmargin=6pt,innerbottommargin=6pt]{mybox}

\usepackage{paracol}
\usepackage{tcolorbox}
\tcbset{colback=mycolor2!5,colframe=mycolor,
fonttitle=\bfseries, float=htb}
\newtcolorbox[blend into=figures]{boxfigure}[3][]
{ float*=ht,width=\textwidth,lower separated=false, center upper,
title={#2},label= fig:#3,#1}
\newtcolorbox[blend into=figures]{smallboxfigure}[3][]
{float=ht,lower separated=false, blend before title=colon hang,
title={#2}, label= fig:#3 ,#1}
\newtcolorbox{smallbox}[3][]
{float=ht,lower separated=false, blend before title=colon hang,
title={#2}, label= fig:#3 ,#1}
\newtcolorbox[blend into=tables]{smallboxtable}[3][]
{float=t,lower separated=false, blend before title=colon hang,
title={#2}, label= table:#3 ,#1}
\newtcolorbox[blend into=tables]{bigboxtable}[3][]
{float*=t,lower separated=false, blend before title=colon hang, width = 2\linewidth,
title={#2}, label= table:#3 ,#1}
\newcolumntype{Z}{|>{\centering\arraybackslash}X}
\usepackage{enumerate}
\hypersetup{
	bookmarksnumbered,
	pdfstartview={FitH},
	citecolor={darkgreen},
	linkcolor={darkred},
	urlcolor={darkblue},
	pdfpagemode={UseOutlines}}
\definecolor{darkgreen}{RGB}{50,190,50}
\definecolor{darkblue}{RGB}{0,0,190}
\definecolor{darkred}{RGB}{238,0,0}
\usepackage{soul}

\newcommand{\be}{\begin{equation}}
\newcommand{\ee}{\end{equation}}
\newcommand{\ben}{\begin{equation*}}
\newcommand{\een}{\end{equation*}}
\newcommand{\bea}{\begin{eqnarray}}
\newcommand{\eea}{\end{eqnarray}}

\newcommand{\tr}{\textnormal{Tr}}

\newcommand{\half}{\mbox{$\textstyle \frac{1}{2}$}}

\newcommand{\ket}[1]{\ensuremath{\left|\right.\!{#1}\!\left.\right\rangle}}

\newcommand{\braket}[2]{\ensuremath{\langle{#1}|{#2}\rangle}}
\newcommand{\ketbra}[2]{\ensuremath{|{#1}\rangle\langle{#2}|}}

\newcommand{\djj}{d\kern-0.4em\char"16\kern-0.1em}

\newcolumntype{s}{>{\hsize=.6\hsize}X}

\newcommand{\nh}[1]{\textcolor{black}{#1}}

\begin{document}

\title{Entangled ripples and twists of light: Radial and azimuthal Laguerre-Gaussian mode entanglement}
\author{Natalia Herrera Valencia}
\email[Email address: ]{nah2@hw.ac.uk}
    \affiliation{Institute of Photonics and Quantum Sciences, Heriot-Watt University, Edinburgh, UK}

\author{Vatshal Srivastav}
    \affiliation{Institute of Photonics and Quantum Sciences, Heriot-Watt University, Edinburgh, UK}
    
\author{Saroch Leedumrongwatthanakun}
    \affiliation{Institute of Photonics and Quantum Sciences, Heriot-Watt University, Edinburgh, UK}
    
\author{Will McCutcheon}
    \affiliation{Institute of Photonics and Quantum Sciences, Heriot-Watt University, Edinburgh, UK}
    
\author{Mehul Malik}
    \email[Email address: ]{m.malik@hw.ac.uk}
    \affiliation{Institute of Photonics and Quantum Sciences, Heriot-Watt University, Edinburgh, UK}
    \affiliation{Institute for Quantum Optics and Quantum Information - IQOQI Vienna, Austrian Academy of Sciences, Vienna, Austria}

\begin{abstract}
It is well known that photons can carry a spatial structure akin to a ``twisted'' or ``rippled'' wavefront. Such structured light fields have sparked significant interest in both classical and quantum physics, with applications ranging from dense communications to light-matter interaction. Harnessing the full advantage of transverse spatial photonic encoding using the Laguerre-Gaussian (LG) basis in the quantum domain requires control over both the azimuthal (twisted) and radial (rippled) components of photons. However, precise measurement of the radial photonic degree-of-freedom has proven to be experimentally challenging primarily due to its transverse amplitude structure. Here we demonstrate the generation and certification of full-field Laguerre-Gaussian entanglement between photons pairs generated by spontaneous parametric down conversion in the telecom regime. By precisely tuning the optical system parameters for state generation and collection, and adopting recently developed techniques for precise spatial mode measurement, we are able to certify fidelities up to 85\% and entanglement dimensionalities up to 26 in a 43-dimensional radial and azimuthal LG mode space. Furthermore, we study two-photon quantum correlations between 9 LG mode groups, demonstrating a correlation structure related to mode group order and inter-modal cross-talk. In addition, we show how the noise-robustness of high-dimensional entanglement certification can be significantly increased by using measurements in multiple LG mutually unbiased bases. Our work demonstrates the potential offered by the full spatial structure of the two-photon field for enhancing technologies for quantum information processing and communication.
\end{abstract}
\maketitle

\section{Introduction}
\label{sec:Intro}

The spatial structure of light allows for the study of a variety of complex phenomena in the classical and quantum regimes~\cite{Forbes2021,Fabre2019}. In this context, the Laguerre-Gaussian (LG) modes of light have emerged as a popular choice of basis. In addition to their aesthetic properties, LG modes form a complete and orthonormal basis of optical modes that are solutions of the paraxial wave equation in cylindrical coordinates~\cite{siegman1986lasers}. As such, they are of inherent interest in many areas of physics featuring cylindrical or circular symmetries. The applications of LG modes extend over research fields as diverse as quantum information \cite{Krenn2017}, matter waves \cite{McMorran:in,FrankeArnold:2017kd}, gravitational waves \cite{BialynickiBirula:2016ju}, and classical telecommunications \cite{Willner:2017jo}, where their relation to the quantisation of orbital-angular-momentum (OAM)~\cite{Allen:1992by}, their unbounded Hilbert space, and their propagation properties have made them of key interest~\cite{Franke-Arnold2008a}.

An LG mode is composed of an azimuthal component given by a twisted helical wavefront of the form $e^{i\ell\phi}$, with the azimuthal index $\ell$ corresponding to the quantised orbital angular momentum (OAM) of photons, and a radial component characterised by the radial index $p$ \cite{Malik:2014ht}. Control over the radial and azimuthal components of light has been key for enhancing classical communication protocols based on mode-division-multiplexing schemes \cite{Zhao:2015hy,Trichili2016,Fontaine2019} and controlling the propagation of light through complex media, with applications ranging from imaging through multi-mode fibres~\cite{Cizmar2012} to the development of programmable optical circuits~\cite{Matthes2019,Leedumrongwatthanakun:2019wt}. Furthermore, access to the complete transverse spatial degree-of-freedom of photons is necessary for harnessing the advantages of high-dimensional encoding for boosting quantum communications with higher capacities~\cite{Mirhosseini:2015fy} and enabling noise-resistant entanglement distribution~\cite{Ecker:2019vx,Zhu2021}.

The capability of performing precise measurements of the radial and azimuthal components of single photons (and their coherent superpositions) is key for their use in quantum information processing~\cite{Krenn2017,Mirhosseini:2015fy}. However, using the full resource of the transverse field is limited by several challenges in accurately measuring full-field Laguerre-Gaussian modes. While multi-outcome measurements in both the azimuthal and radial part of spatial modes have been demonstrated through interferometric techniques~\cite{Gu2018a,Fu2018,Zhou2017,Ionicioiu:2016dr}, the scalability of such implementations is difficult. A promising alternative is the use of multi-plane mode converters~\cite{Morizur2010,Choudhary2018, Fontaine2019,Fickler2020}, but loss and cross-talk present additional challenges. On the other hand, single-outcome measurements in the spatial degree-of-freedom can be performed through multi-plane phase modulations~\cite{Hiekkamaki2019a}, or well-established phase-flattening techniques~\cite{Mair2001}. In the latter, spatial filters composed of spatial light modulators (SLM) and single-mode fibres (SMFs) achieve accurate projective measurements of spatial modes. In this case, limitations arise from the loss associated with the use of SLMs, where it is necessary to implement intensity masking in order to modulate the amplitude of the field~\cite{Arrizon:2007wl}. In addition, the issue of mode-dependent detection efficiencies presents additional hurdles, especially when coherent superpositions of modes are being measured~\cite{Qassim:2014fp,Bouchard:2018hr}. 

While the precise description of the full transverse structure of light requires both indices ($\ell$ and $p$), the generation and detection of LG modes in the quantum domain initially focused on their azimuthal component~\cite{Mair2001,Franke-Arnold2008a}, with early experiments demonstrating their use in quantum communication and entanglement (please see \cite{Krenn2017} for a review of these). The ``forgotten'' radial quantum number $p$~\cite{Plick2013} has emerged in recent years as the subject of several theoretical investigations~\cite{Miatto:2011cr,Karimi2012,Karimi2014,Plick2015}, with various experimental demonstrations of quantum radial correlations in the transverse field of photons produced through spontaneous parametric down-conversion (SPDC)~\cite{Zhang2014,Zhang2018,Liu2019,DErrico2021}. These studies have shown that although perfect azimuthal correlations can be easily observed due to the conservation of OAM, entanglement in the radial part is only available when the the pump and detected mode waists are finely tuned~\cite{Miatto:2011cr,Salakhutdinov2012}. Other experiments have implemented spatial-mode measurements that only modulated the phase of the LG~modes~\cite{Salakhutdinov2012,Karimi2014b,Krenn:2014jy}, or used phase-only discretisation of the radial space through Walsh functions~\cite{Geelen2013}. However, access to the full modal bandwidth of spatially entangled photons through accurate measurements of radial modes requires both amplitude and phase-sensitive detection. This becomes particularly clear when implementing projections onto coherent superposition states, which are of key importance when studying entanglement. 

Here we demonstrate the generation and measurement of the full transverse spatial field of a two-photon state generated through spontaneous parametric down-conversion (SPDC) in the telecom regime, certifying up to 26-dimensional radial and azimuthal Laguerre-Gaussian mode entanglement with a dimensionality of 43. Accurate spatial state projections onto states of the LG basis (and any superposition of these) are performed with an ``intensity-flattening'' technique that we have recently demonstrated in both the classical and quantum regime~\cite{Bouchard:2018hr,HerreraValencia2020}. Our measurement scheme optimises correlations in the radial component through the fine adjustment of the mode sizes in both the generation and detection systems, allowing us to certify entanglement between transverse spatial modes of light spanning over 21 LG mode groups. In addition, our high-dimensional entangled states are generated at 1550~nm, making it possible to interface them with optical fibers at extremely low loss for quantum information schemes based on space-division multiplexing~\cite{Xavier2020}, and enabling the realisation of very bright entangled sources~\cite{Pickston2021} for multi-photon experiments. 

\section{Theory}
\label{sec:Theory}
\begin{figure*}[t]
    \centering
    \includegraphics[width=0.9\linewidth]{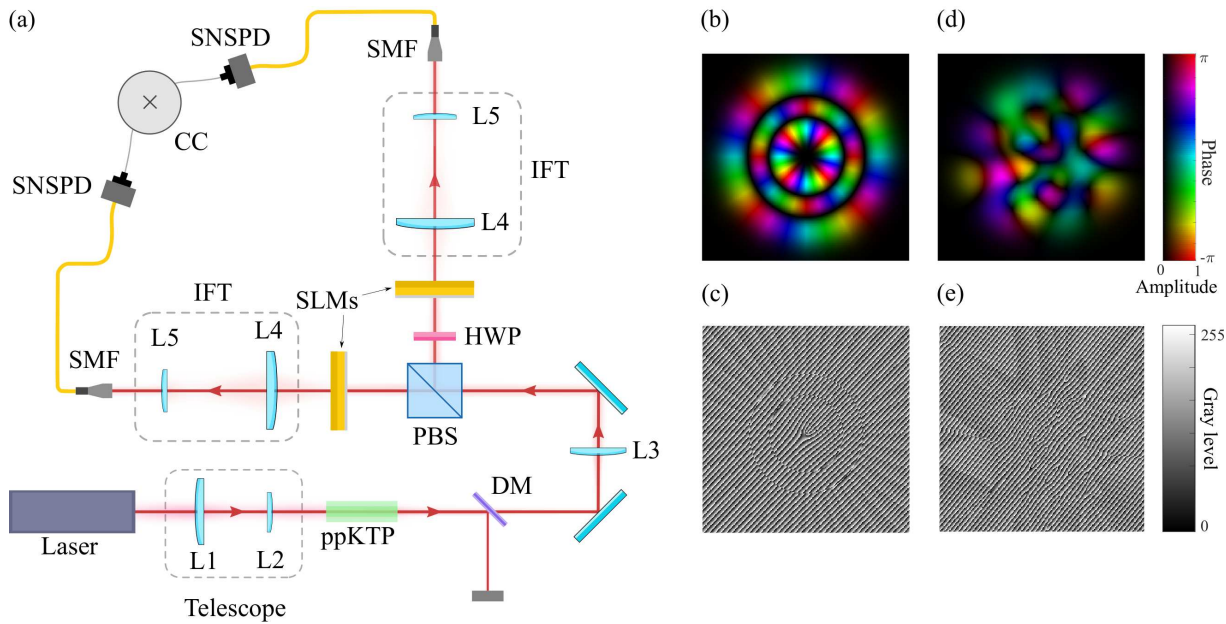}
    \caption{\textbf{Experimental Setup:} a) A 775-nm Ti:Sapphire pulsed laser with a beam waist tailored by a telescope system pumps a non-linear ppKTP crystal and generates pairs of entangled photons at 1550~nm through Type-II spontaneous-parametric-down-conversion (SPDC). Accurate projective measurements in the Laguerre-Gauss basis and any of its mutually unbiased basis are performed with spatial light modulators (SLMs), intensity-flattening telescopes (IFT), and single-mode fibres (SMF). Correlations in the radial and azimuthal LG mode components of the transverse spatial field are obtained by measuring coincidence counts between pairs of photons using two superconducting nanowire detectors (SNSPD) connected to a coincidence counting logic (CC). b) Representation of the complex amplitude describing the full-field mode $LG_{2}^{4}$ and d) the LG MUB mode corresponding to state $\ket{\tilde{1}_1}$ from the second mutually unbiased basis w.~r.~t.~LG basis, and composed of a coherent superposition of 43 LG modes with $p=0,...,4$ and $\ell=-8,...,7$. The color scale corresponds to the phase of the mode and the brightness to the absolute value of its amplitude. c) and e) Examples of corresponding computer generated holograms (Type 1 complex amplitude modulation given in~\cite{Arrizon:2007wl}) displayed on the SLMs to modulate the phase and amplitude of the LG and LG MUB modes in order to measure their spatial mode content.}
    \label{fig:expSetup}
\end{figure*}

To understand entanglement in the azimuthal and radial components of the transverse spatial field of a pair of photons created via SPDC, let us write the two-photon state in the LG spatial mode basis:
\be
\ket{\Psi_{\textrm{LG}}} = \sum_{\ell_i,p_i,\ell_s,p_s} C_{p_sp_i}^{\ell_s\ell_i} \ket{\ell_s p_s}\ket{\ell_i p_i }.
\label{eq:SPDC}
\ee 
Here, the subscripts $s$ and $i$ refer to the signal and idler photon respectively. For simplicity, we use the expressions $\ket{\ell_{s} p_{s}}$ and $\ket{\ell_{i} p_{i}}$ in Eq.~\eqref{eq:SPDC} to refer to a single photon in modes $LG^{\ell_s}_{p_s}$ and $LG^{\ell_i}_{p_i}$, respectively. The coefficient $|C_{p_sp_i}^{\ell_s\ell_i}|^2$ indicates the joint probability of finding a signal photon in mode $LG^{\ell_s}_{p_s}$ and an idler photon in mode $LG^{\ell_i}_{p_i}$. While these coefficients are primarily determined by the SPDC process, they also depend on the collection optics involved in the measurement scheme. An ideal LG entangled state exhibits perfect two-photon anti-correlations in azimuthal modes $(\ell_s=-\ell_i)$ and perfect correlations in radial modes $(p_s=p_i)$. However, due a variety of reasons that can be attributed to measurement imperfections and the optics used, cross-talk between modes can appear that reduces the quality of the measured state. Below, we briefly discuss the origin of radial mode cross-talk.

An LG mode in the transverse momentum space can be explicitly written in cylindrical coordinates as \cite{Miatto:2011cr}:
\bea
LG_p^\ell(\rho,\phi) &=& \sqrt{\frac{w^2p!}{2\pi(p+|\ell|)!}}\left(\frac{\rho w}{\sqrt{2}}\right)^{|\ell|} \\
&\times& \exp\left(\frac{-\rho^2 w^2}{4}\right)L_p^{|\ell|}\left(\frac{\rho^2w^2}{2}\right)\exp\left[i\ell\left(\phi\right)\right], \nonumber   
\label{eq:LGmode}
\eea
where $w$ is the beam waist in the transverse position space (we have assumed $z=0$), and $L_p^{|\ell|}(\cdot)$ is the associated Laguerre polynomial. In order to understand the entanglement certification method used here, it is also important to define bases that are mutually unbiased with respect to the LG basis (LG MUB). For prime dimensions $d$, these can be calculated by following the construction \cite{Wootters1989}: 
\be
\ket{\tilde{j}_r} = \frac{1}{\sqrt{d}}\sum_{m=0}^{d-1}\varepsilon^{jm+rm^2}\ket{m}
\label{eq:WFbasis}
\ee
where $\{\ket{m}\}_m$ denotes the standard LG basis, $\varepsilon = \exp\left({\frac{2\pi i}{d}}\right)$ is the principal complex d-$th$ root of unity, $r\in\{0,\dots,d-1\}$ labels the chosen LG MUB, and $j\in\{0,\dots,d-1\}$ labels the basis elements. Figs.~\ref{fig:expSetup}.b,d show examples of LG and LG MUB modes carrying both azimuthal and radial components (``twists and ripples'').

The coefficients $C_{p_sp_i}^{\ell_s\ell_i}$ in Eq.~\eqref{eq:SPDC} can be obtained by projecting the measured biphoton state onto LG mode operators for the signal and idler photons. This is calculated by taking the overlap integral between the measured state and the corresponding signal and idler mode functions over the wave vector space:
\bea
C_{p_sp_i}^{l_sl_i} &=& \braket{\ell_{s} p_{s}\: \ell_{i} p_{i}}{\Psi_{\textrm{LG}}}\label{eq:overlap} \\
&=& \int \int d^3k_sd^3k_i \Phi(\mathbf{k}_s,\mathbf{k}_i)[LG_{p_s}^{\ell_s}(\mathbf{k}_s)]^*[LG_{p_s}^{\ell_i}(\mathbf{k}_i)]^*, \nonumber
\eea
where $\Phi(\mathbf{k}_s,\mathbf{k}_i)$ is the measured joint transverse momentum amplitude (JTMA) of the biphoton state~\cite{Srivastav2021}. The JTMA is a function determined by the spatial profile of the pump, the phase matching conditions in SPDC~\cite{Schneeloch:2016ch,Walborn:2010dd,Miatto:2011cr,Srivastav2021}, as well as the collection optics. The JTMA describes the measured correlations in the transverse-momentum degree-of-freedom of the generated photons.

The result of the overlap integral in Eq.~\eqref{eq:overlap} has been shown to depend on the real space ratio between the pump and the down-converted signal and idler mode waists, $\gamma = w_p/w_{s,i}$~\cite{Miatto:2011cr}. The mode waists $w_i$ and $w_s$ are equivalent to the ``collected'' mode waists obtained at the crystal. The azimuthal part of the overlap integral enforces the conservation of OAM, which for the case of a Gaussian ($\ell=0$) pump gives perfect two-photon correlations between azimuthal modes with indices $\ell_s = -\ell_i$. However, correlations between the radial modes are only perfect in the limit of an infinite pump waist, which breaks down as the ratio $\gamma \to 1$~\cite{Miatto:2011cr,Salakhutdinov2012}. In fact, in any realistic experimental setup with finite-sized apertures and optics, the pump waist is limited to a large extent. Harnessing the resource of entanglement in the full transverse field thus requires one to increase the ratio $\gamma$ by optimising the pump and collection mode waists appropriately in order to maximise the correlations between Laguerre-Gaussian modes with non-zero radial components ($p_s$, $p_i>0$). 

\section{Experiment}
As shown in Fig.~\ref{fig:expSetup}a, we use a Ti:Sapphire femtosecond pulsed laser ($\lambda_p = 775$nm, 500mW average power) to pump a non-linear ppKTP crystal (1mm $\times$ 2mm $\times$ 5mm) in order to generate pairs of spatially entangled photons at 1550~nm through Type-II spontaneous-parametric-down-conversion (SPDC). A telescope composed of lenses L1 and L2 is used to set the $1/e^2$ radius of the pump to be $w_p = 450\mu$m at the crystal (this is the Gaussian pump mode waist considered in the ratio $\gamma$). After filtering out the pump with a dichroic mirror (DM), the signal and idler photons are separated using a polarising beam splitter (PBS), and sent to spatial light modulators (SLMs) that are placed in the Fourier plane of the crystal via a 250~mm lens (L3). Note that reflection on the PBS flips the sign of the idler mode azimuthal index, converting azimuthal anti-correlations into correlations. To perform accurate projective measurements over the complete set of LG modes spanning the field (and any superposition of these), the spatially varying amplitude and phase of the incoming photons are modulated by displaying computer-generated holograms (CGH) on the SLMs~\cite{Arrizon:2007wl}. Fig.~\ref{fig:expSetup}b depicts an example full-field mode ($LG_{2}^{4}$) containing both radial ($p=2$) and azimuthal ($\ell=4$) components, with Fig.~\ref{fig:expSetup}c showing the corresponding CGH used to measure it. Figs.~\ref{fig:expSetup}d and e depict a full-field mode and its corresponding CGH from the second ($r=1$ in Eq.\eqref{eq:WFbasis}) 43-dimensional LG MUB containing a coherent superposition of modes with $p=0,...,4$ and $\ell=-8,...,7$.

The CGH displayed on an SLM is used for filtering a particular spatial mode such that it effectively couples to a single-mode fibre (SMF). This method relies on the orthogonality of the LG modes, where a given $LG_p^\ell$ is measured by ``flattening'' its complex amplitude with a CGH of its complex conjugate $[LG_p^\ell]^*$. However, the use of an SMF in this measurement scheme introduces an additional Gaussian factor that results in undesired cross-talk between modes with different indices, which is particularly increased for radial modes~\cite{Qassim:2014fp,Bouchard:2018hr}. In order to minimise this cross-talk, the orthogonality between the input and projected mode is maintained through the use of a so-called intensity-flattening telescope (IFT, lenses L4 and L5) placed between the SLM and the SMF~\cite{Bouchard:2018hr}. The IFTs in our system effectively magnify the back-propagated collection modes at the SLM planes by a factor of $3.3$ to mitigate the effects of the SMF Gaussian component. In addition, the IFTs allow us to tailor the collection modes to the size of the quantum spiral bandwidth of the generated state~\cite{Torres:2003cy}, as determined by the extent of the JTMA~\cite{HerreraValencia2020,Srivastav2021}. Note also that increasing the size of the collection modes at the SLMs is equivalent to reducing the size of the collection mode waists $w_{s,i}$ at the crystal, which increases the ratio $\gamma_{exp}$ to a value of 5.26, in turn reducing radial mode cross-talk between signal and idler photons~\cite{Miatto:2011cr}.

\section{Results}

\begin{figure}[t]
    \centering
    \includegraphics[width=1\linewidth]{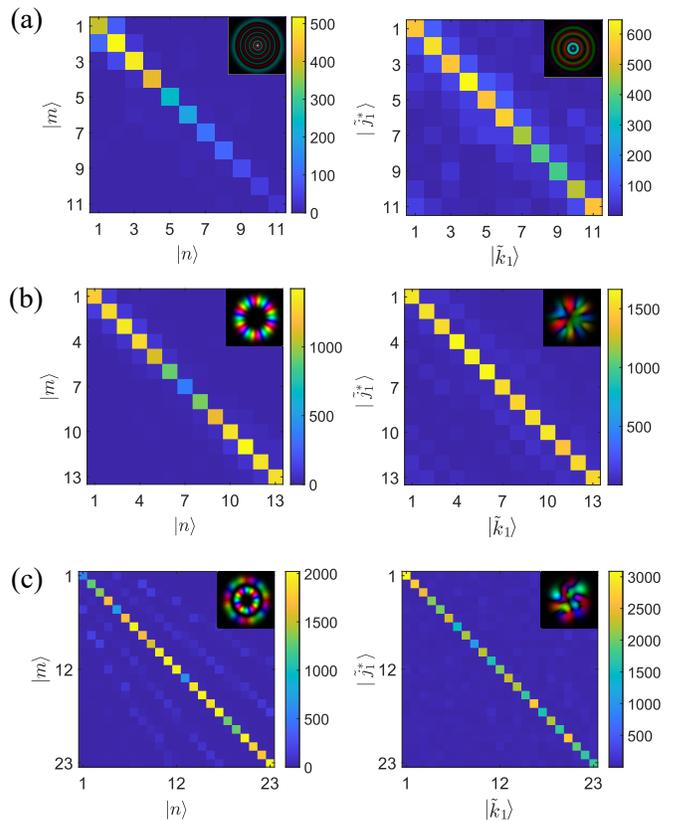}
    \caption{\textbf{Two-photon correlations in azimuthal and radial LG modes.} Two-photon coincidence counts showing radial (a), azimuthal (b), and full-field (c) correlations in the standard LG basis of each defined subspace $\{\ket{m},\ket{n}\}_{m,n}$ (left), and in its second mutually unbiased basis $\{\ket{\tilde{j}_1},\ket{\tilde{k}_1}\}_{j,k}$ (right). The inset on each figure illustrates the highest order mode measured in that particular basis. Measuring the correlation matrices for all mutually unbiased bases (MUBs) of each high-dimensional space, we obtain fidelities to the maximally entangled state shown in the fourth column of Table~\ref{table:Flist}, allowing us to certify entanglement dimensionalities of $d_{\text{ent}}=7$ in the 11-dimensional radial subspace, $d_{\text{ent}}=11$ in the 13-dimensional azimuthal subspace, and $d_{\text{ent}}=18$ in the 23-dimensional full-field subspace.}  
    \label{fig:correlations}
\end{figure}

\begin{figure*}[ht!]
    \centering
    \includegraphics[width=\linewidth]{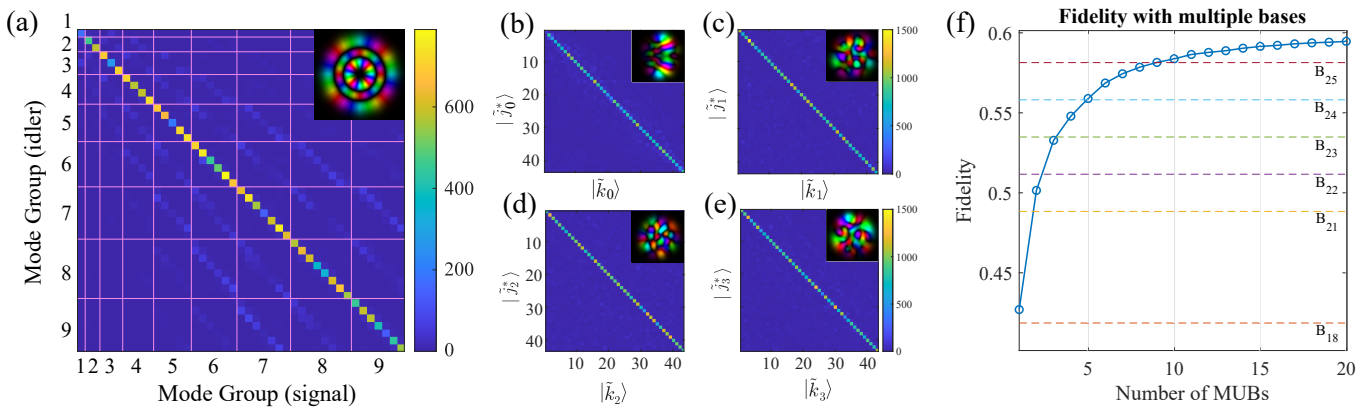}
    \caption{\textbf{Full LG-mode entanglement in a 43-dimensional subspace.} (a) Two-photon coincidence counts showing correlations in the standard LG basis of radial and azimuthal modes belonging to 9 different mode groups (indicated by the pink lines). (b-e) Two-photon coincidence counts showing correlations in the first 4 mutually unbiased bases (LG MUBs) with respect to the standard basis. Correlations in 21 mutually unbiased bases allow us to lower bound the fidelity of our state to a maximally entangled state, and certify an entanglement dimensionality of $d_{\text{ent}}=26$. The advantage of using measurements in more LG MUBs is shown in (f), where the estimated fidelity allows us to violate higher dimensionality bounds ($B_{d_{\text{ent}}-1}$), thus allowing us to certify higher entanglement dimensionality as the number of MUBs used increases.}
    \label{fig:FullField}
\end{figure*}

We measure two-photon correlations in the azimuthal and radial components of the transverse field with measurements in the LG (standard, or computational) basis using CGHs for radial modes ($\ell=0$, $p\geq0$), azimuthal modes ($\ell\in\mathbb{Z}$, $p=0$), and full-field modes ($\ell\in\mathbb{Z}$, $p\geq0$). With the ability of projecting into any given superposition of spatial modes, we also perform measurements in all mutually unbiased bases with respect to the LG basis (LG MUBs) following the construction given in~\cite{Wootters1989}. Measurements in the complete set of LG MUBs allow us to determine the exact fidelity $F(\rho,\ket{\Phi^+}) = \tr(\ketbra{\Phi^+}{\Phi^+}\rho)$ of our experimentally measured state $\rho$ to the maximally entangled state $\ket{\Phi^+} = \frac{1}{\sqrt{d}}\sum_{\ell,p}\ket{\ell\, p}_s\ket{\ell\, p}_i$, where $d$ is the entanglement dimensionality and $d\times d$ is the number of entangled modes in the state~\cite{Bavaresco:2018gw}. By comparing our measured fidelity to fidelity bounds calculated from the Schmidt coefficients of the target state \ket{\Phi^+}, we can certify the entanglement dimensionality of our experimental state.

We construct radial mode bases in dimension $d=11$~($p=0,\dots,10$), azimuthal mode bases in $d=13$~($\ell=-6,\dots,6)$, and full-field LG mode bases in $d=23$~($\ell=-6,\dots,5$ and $p=0,1,2$). Since a complete set of $d+1$ mutually unbiased bases can only be constructed for prime dimensions $d$, we use an asymmetrical azimuthal bandwidth to build the full-field LG MUBs in a prime dimension. 

As shown in Fig.~\ref{fig:correlations}, our measurements result in strong diagonal two-photon correlations over the complete set of the LG and LG MUBs of the radial, azimuthal and full-field mode spaces. The off-diagonal correlations in the standard radial basis (Fig.~\ref{fig:correlations}a) correspond to the expected cross-talk originating from the overlap integral in Eq.~\ref{eq:overlap} with our $\gamma_{\text{exp}} = 5.26$, with additional cross-talk due to experimental imperfections such as alignment. The cross-talk between adjacent azimuthal modes, i.e. $\ell_s = \ell_i \pm 1$ observed in Fig.~\ref{fig:correlations}b can be attributed to experimental imperfections such as misalignment and finite SLM resolution. These two effects can also be observed in the full-field correlations shown in Fig.~\ref{fig:correlations}c. The cross-talk in the LG MUBs (second column of Fig.~\ref{fig:correlations}) arises from the uneven distribution of the diagonal correlations in the standard basis. This probability distribution is determined by the quantum spiral bandwidth of the SPDC state, which dictates a higher coincidence probability for lower order LG modes~\cite{Miatto:2011cr}, and a mode-dependent coupling efficiency that originates from the use of SMFs~\cite{Qassim:2014fp,Bouchard:2018hr}. Optimising the parameter $\gamma$ increases the available entanglement by flattening the quantum spiral bandwidth, allowing us to access higher-order correlations and reducing the cross-talk in the LG MUB measurements. 
 
Even though we can optimise the correlations over the transverse spatial field of the biphoton state through the manipulation of the pump and collection mode waists, a ``flat'' spiral bandwidth can only be achieved with an infinite pump waist. Since the SPDC state is a non-maximally entangled state, one can also characterise its spatial correlations through a tilted-basis witness that uses the fidelity to a general, non-maximally entangled target state $F(\rho,\Phi)$~\cite{Bavaresco:2018gw}. The performance of this tilted witness improves with an appropriate choice of the target state, with fidelity ideally approaching unity. An informed guess for a good target state can be made from the measured correlations in the standard basis, which then informs the construction of appropriate tilted-bases, in manner similar to LG MUBs. In this process, it is important to correct for the effects of mode-dependent loss in the coincidence counts of the LG basis, as demonstrated in~\cite{Bavaresco:2018gw}.
 
Fidelities to the maximally entangled state (\ket{\Phi^+}) and a non-maximally entangled target state (\ket{\Phi}) in each mode space are shown in Table~\ref{table:Flist}. Errors in the fidelities are calculated via Monte-Carlo simulation of the experiment that propagates statistical Poisson error associated to the photon count rates. With an appropriate choice of target state, is possible to obtain higher fidelities $F(\rho,\Phi)$ compared to the ones calculated with respect to the maximally entangled state $F(\rho,\Phi^+)$. Nevertheless, the fidelity bounds for certifying entanglement dimensionality become harder to violate for the tilted-witness, resulting in the same or lower certified dimensionality. The entanglement dimensionalities shown in Table~\ref{table:Flist} are thus calculated using the fidelity $F(\rho,\Phi^+)$ of our measured state to a maximally entangled target state~\cite{Bavaresco:2018gw}.

\begin{bigboxtable}[floatplacement=t]{Measured fidelities $F(\rho,\ket{\Phi^+})$ and $F(\rho,\ket{\Phi})$ of experimental LG-entangled states ($\rho$) to the maximally ($\ket{\Phi^+}$) and non-maximally ($\ket{\Phi}$) entangled target states}{Flist}
\begin{center}
\begin{tabularx}{\textwidth}{p{0.15\textwidth}p{0.15\textwidth}p{0.15\textwidth}p{0.075\textwidth}p{0.075\textwidth}p{0.20\textwidth}p{0.20\textwidth}}
\hline
    Type & $\ell$ & $p$ & $d$ & $d_{\mathrm{ent}}$ & $F(\rho,\Phi^+)$ & $F(\rho,\Phi)$\\
\hline
Radial & 0 & $0,\dots,10$ & 11 & 7& $61.8 \pm 0.5$ \% & $69.0 \pm 1.0$ \% \\
Azimuthal & $-6,\dots,6$ & 0 & 13& 11& $83.3 \pm 0.3$ \%  & $85.7\pm 0.6$ \%\\
Full Field & $-6,\dots,5$ & $0,1,2$ & 23& 18& $75.1 \pm 0.1$ \% & $75.5 \pm 0.1$ \%\\
Full Field$^\dagger$ & $-8,\dots,7$ & $0,\dots,4$ & 43& 26& $59.5 \pm 0.9$ \% & $\quad$ -- \\
\end{tabularx}
\end{center}
\small{The third column lists the entanglement dimensionality $d_{\mathrm{ent}}$ certified using the fidelity to the maximally entangled state $F(\rho,\Phi^+)$.}\\
\footnotesize{$\dagger$ This fidelity value corresponds to a lower bound to the fidelity obtained from measurements in 21 of the 44 MUBs of the 43-dimensional space.}
\end{bigboxtable}

To demonstrate the potential of our technique for accessing very high-dimensional LG entanglement, we use azimuthal and radial modes to construct a full-field standard LG basis with dimension $d=43$, composed of LG modes with $\ell=-8,\dots,7$ and $p=0,\dots,4$. This basis contains states with mode group order ($MG = 2p+|l|+1$) up to 9. The LG mode groups consist of modes that experience the same Gouy phase, and are of immense interest in the study of multi-mode waveguides. As shown in Fig.~\ref{fig:FullField}a, the cross-talk between modes in the standard basis has a particular structure. When examining them with respect to their respective mode groups (pink lines), we can see that very little cross-talk is present within the same mode group. However, alignment imperfections leading to correlations between azimuthal modes with $\ell_s \neq \ell_i$ for a given $p$ result in cross-talk between adjacent mode groups $MG_i=MG_s\pm 1$. On the other hand, radial mode correlations between modes with $p_s\neq p_i$ manifest as cross-talk that skips a mode group $MG_i=MG_s\pm 2$. This structure can be understood as a natural consequence of the way mode groups are defined, with a factor of two appearing before the index $p$.

Instead of estimating the exact fidelity through measurements in all LG MUBs for a given dimension, the certification of high-dimensional entanglement is possible by lower bounding the fidelity to a given target entangled state with measurements in at least two MUBs. \nh{It has been theoretically shown that as the number of bases used for lower bounding the fidelity is increased, the detrimental effect that noise (manifested as cross-talk in the correlations) has on the fidelity bound is reduced, resulting in the certification of entanglement dimensionalities under a higher noise tolerance~\cite{Bavaresco:2018gw}}
This is especially advantageous in scenarios where there is a lot of noise present to begin with. Here we experimentally demonstrate this noise advantage, as can be clearly seen in Fig.~\ref{fig:FullField}f. Increasing the number of LG MUBs used for lower bounding the fidelity from 2 to 21 results in an improvement in the fidelity bound from $\tilde{F}(\rho,\Phi^+) = 44.6\pm 0.9\%$ to a value of $59.5\pm 0.9\%$. This improvement in fidelity corresponds to an increased certified entanglement dimensionality from $d_{\text{ent}}^{\text{Full}}=20$ to $26$. The results show that the use of additional MUBs offers a significant advantage for entanglement certification in noisy regimes.

\section{Conclusion}
We have demonstrated high-dimensional entanglement in the telecom regime between two photons in their full-field Laguerre-Gaussian (LG) spatial mode basis consisting of radial and azimuthal components. We are able to harness the complete resource of LG mode entanglement with measurement settings tailored to the spatial distribution of the measured two-photon state, and an intensity-flattening technique that ensures accurate state projections onto any given spatial mode, while minimising radial mode cross-talk. Careful control over the mode waists of the pump and collected photons increases the correlation strength within the radial and azimuthal components of the two-photon field and allows for the certification of entanglement dimensionalities up to 26. By measuring correlations in a 43-dimensional set of LG modes and its mutually unbiased bases (LG MUBs), we are able to characterise high-dimensional entanglement in modes spanning nine LG mode groups, and observe how two-photon inter-modal cross-talk follows a structure related to the mode group orders. In addition, we demonstrate clearly how measurements in additional LG MUBs enables one to certify high-dimensional entanglement in a noise-robust manner. Our techniques are significant for the emerging field of high-dimensional quantum information and will prove beneficial for quantum technologies harnessing Laguerre-Gaussian modes of light.

Note: We have recently become aware of a related work~\cite{DErrico2021} that characterised LG radial mode entanglement through full quantum state tomography in state spaces with local dimension 4. 

 \begin{acknowledgements}
  This work was made possible by financial support from the QuantERA ERA-NET Co-fund (FWF Project I3773-N36) and the UK Engineering and Physical Sciences Research Council (EPSRC) (EP/P024114/1).
 \end{acknowledgements}


\bibliographystyle{apsrev4-1fixed_with_article_titles_full_names}
\bibliography{refs2}

\clearpage
\onecolumngrid
\appendix
\section*{Appendix}
\renewcommand{\thesubsection}{A.\Roman{subsection}}
\renewcommand{\thesection}{}
\setcounter{equation}{0}
\numberwithin{equation}{section}
\renewcommand{\theequation}{A.\arabic{equation}}
\setcounter{figure}{0}
\renewcommand{\thefigure}{A.\arabic{figure}}
\renewcommand{\theHfigure}{A.\arabic{figure}}

We have demonstrated the generation and measurement of high-dimensional full-field Laguerre-Gaussian entanglement between two pairs of photons generated by spontaneous parametric down conversion at telecom wavelength. In this Supplementary Information, we provide additional information on the use of the Joint Transverse Momentum Amplitude to optimise the correlations in the Laguerre-Gaussian basis. Furthermore, we discuss the importance of implementing amplitude and phase modulation in the precise measurement of full-transverse spatial field modes, and provide details on the mode-dependent efficiency correction implemented on the certification of entanglement with our tilted fidelity witness. Additionally, we provide further details of the full-field modes used in our experiment to further illustrate the richness of the full spatial structure of light. 

\subsection{Joint Transverse Momentum Amplitude}

Let us consider the \textit{joint-transverse-momentum-amplitude} (JTMA)~\cite{Srivastav2021}, describing the correlations in the momentum space of the biphoton state generated through the Type-II spontaneous parametric down-conversion process (SPDC). This function can be well approximated in the degenerate case by~\cite{Schneeloch:2016ch}:
\be
\Phi(\bm{k}_s, \bm{k}_i)= \mathcal{N}_1\underbrace{\exp\bigg( \frac{-|\bm{k}_s+\bm{k}_i|^2}{2\sigma_P^2}\bigg)}_{\textrm{Pump Profile}}\times \underbrace{\text{ sinc}\bigg(\frac{1}{\sigma_S^2}|\bm{k}_s - \bm{k}_i|^2\bigg)}_{\textrm{Phase-Matching Condition}},
\label{eq:JTMAformula}
\ee
where $\mathcal{N}_1$ is the normalization constant, and $\bm{k}_s (\bm{k}_i)$ is the transverse-momentum vector for the signal (idler) photon. While the first term of Eq.~\ref{eq:JTMAformula} arises from the Gaussian profile the transverse pump momentum, the sinc function describes the phase-matching condition that the SPDC process enforces along the crystal's $z$-component. The momentum correlations dictated by the JTMA are determined by the width of the pump's profile $\sigma_P$, and the phase-matching width $\sigma_S$. The former can be understood as the strength of the correlations in the momentum space, and physically depends on the pump's waist. The latter gives an idea of the extent of entanglement along the transverse-momentum space and physically depends on the crystal's length and pump's wavelength. 

While the JTMA determines the correlations of the generated state, the measured entangled state also depends on the detection scheme. As discussed in the main text, the state projection technique of spatial-mode filtering through spatial light modulators and coupling to single mode fibres (SMF) determines the ``collected" mode waists $w_{s,i}$. On the momentum space at the crystal, the collection mode takes the form of the back-propagated SMF Gaussian mode: $\mathcal{C}(\bm{k})=(\sqrt{\pi}\sigma^{(x)}_C)^{-\half}\exp\{- \tfrac{|\bm{k}|^2}{2\sigma_C^2}\}$, characterised by collection bandwidth:
$\sigma_C=\tfrac{2\pi \sigma_C^{(x)}}{f \lambda}$, where $f$ is the focal length of the lens, and the collection width in the real space $\sigma_C^{(x)}$ relates to the collection beam waists as  $w_{s_i} = \sqrt{2}\sigma_C^{(x)}$.

The choice of $\sigma_C$ (relative to $\sigma_S$ and $\sigma_P$) limits the quantum spiral bandwidth, and the quality of the correlations between spatial modes. We can thus consider the \textit{collected} bi-photon JTMA
\bea
G(\bm{k}_s, \bm{k}_i)&=&\mathcal{C}(\bm{k}_s)\times\mathcal{C}(\bm{k}_i)
\times F(\bm{k}_s, \bm{k}_i) \\
&=&\tfrac{\mathcal{N}_1}{\sqrt{\pi}\sigma_C}\underbrace{\exp\bigg(- \frac{|\bm{k}_s|^2}{2\sigma_C^2}\bigg)
\exp\bigg(- \frac{|\bm{k}_i|^2}{2\sigma_C^2}\bigg)}_{\text{Collection widths}}
\underbrace{\exp\bigg( \frac{-|\bm{k}_s+\bm{k}_i|^2}{2\sigma_P^2}\bigg)}_{\text{Correlations strength}} \times \underbrace{\text{sinc}\bigg(\frac{1}{\sigma_S^2}|\bm{k}_s - \bm{k}_i|^2\bigg)}_{\text{Generation width}} \nonumber .
\label{eq:collectedJTMA}
\eea

The knowledge of the collected JTMA provides an idea of the distribution of the correlations of the down-converted photon on the SLM plane, and allows us to optimise the probability of measuring higher-order correlations by making sure that all of the modes we are projecting on match the width of the JTMA defined by $\sigma_C$. To do so, we adjust the size of the holograms displayed on the SLM, and use the intensity-flattening telescope to magnify the Gaussian collection mode envelope on the plane of the SLM by a factor of $3.3$, which effectively increases the available quantum spiral bandwidth. Notice that ensuring these conditions on the Fourier plane of the crystal (SLM plane) is equivalent to increasing the factor $\gamma$ in the real space, therefore harnessing higher-dimensional entanglement. 

\subsection{The role of amplitude modulation in the precise measurement of complex photonic modes}

In the main text we have stressed the importance of using both amplitude and phase modulation to perform accurate measurements over the full Laguerre-Gaussian modal bandwidth of the transverse spatial field of light. Here we will elaborate on how phase-only measurements hinder the study of LG-entanglement and may lead to fundamental errors in the characterisation of the correlations over the transverse spatial field.

Characterising entanglement through dimensionality witnesses relies on trusting that our devices are performing the correct measurements. In particular, to certify dimensionality with the witness in~\cite{Bavaresco:2018gw}, we rely on the orthogonality of the elements in the standard basis: $|\braket{m}{n}|^2 = \delta_{mn}$, the orthogonality of the elements of a MUB: $|\braket{\tilde{i}}{\tilde{j}}|^2 = \delta_{ij}$ and the mutual unbiasedness between the standard and MUB: $|\braket{m}{\tilde{j}}|^2 = 1/d$ $\forall m$ within the standard basis $\{\ket{m}\}_m$. 

\begin{figure*}[ht!]
    \centering
   \includegraphics[width=0.75\linewidth]{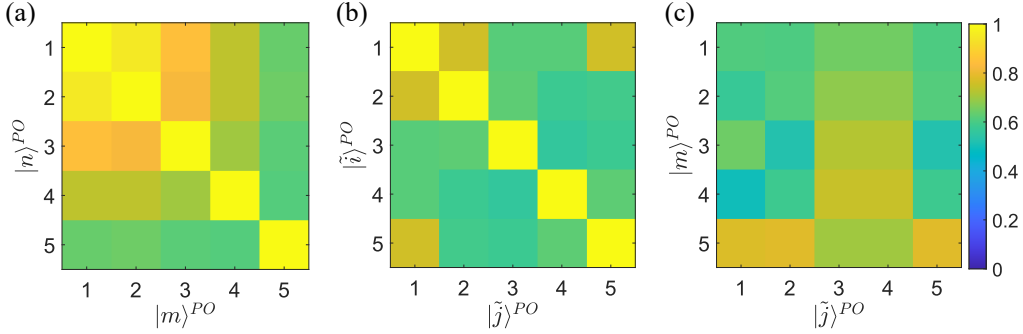}
    \caption{\textbf{Phase-only radial measurements:} (a) Overlaps between phase-only radial modes $\{\ket{m}\}_m^{PO}$ demonstrate that they don't form an orthogonal basis. The same applies for the overlap between the radial MUB modes $\{\ket{\tilde{j}}^{PO}\}_j$ in (b). (c) The overlap between modes of the two different sets $|\braket{m}{\tilde{j}}|^2$ demonstrates that they aren't mutually unbiased, and thus can't be used for certifying entanglement.}
    \label{fig:phaseonly}
\end{figure*}

Let us then consider phase-only radial measurements in a 5-dimensional space $\{\ket{m}\}_m^{PO}$ ($\ell = 0$ and $p = 0,4$), and the set of phase-only modes $\{\ket{\tilde{j}}^{PO}\}_j$ calculated according to Eq.~3 of the main text. The elements of each set are constructed with an amplitude of 1 and a phase given by the argument of the complex amplitude describing the corresponding radial LG or radial LG MUB state. As shown in Fig.\ref{fig:phaseonly} a. and b., the non-zero overlap between different elements of the same set shows that phase-only radial modes can't be used to form orthogonal bases. Furthermore, Fig.~\ref{fig:phaseonly}.c depicts the overlap between the LG radial basis and the radial MUB, demonstrating that the two sets can't be used as mutually unbiased bases.

\subsection{Mode Dependent Efficiency Correction}
\begin{figure*}[ht!]
    \centering
    \includegraphics[width=0.75\linewidth]{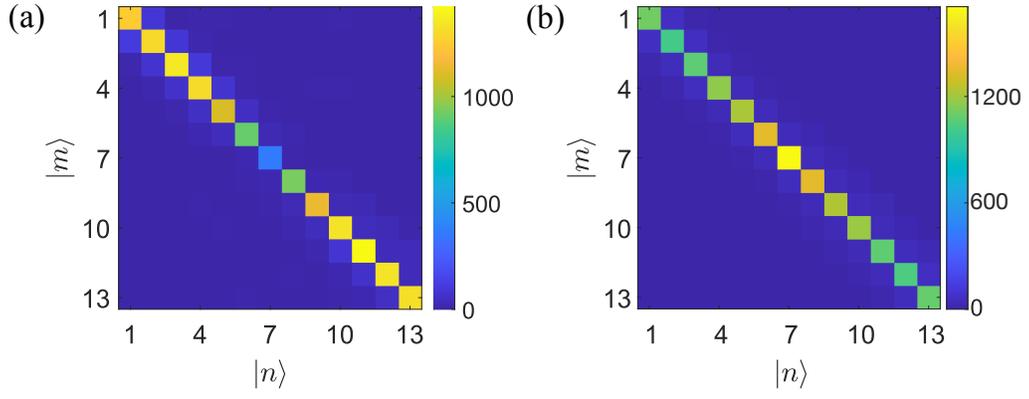}
    \caption{\textbf{Example of mode-dependent loss correction} Two-photon coincidence counts showing correlations in the standard LG basis before (a) and after (b) the correction of the mode-dependent loss.}
    \label{fig:EtaCorrection}
\end{figure*}

To characterise the spatial correlation of a non-maximally entangled state like the one created through SPDC, we lower bound the fidelity to an entangled target state with a tilted-basis witness~\cite{Bavaresco:2018gw}. The performance of this dimensionality witness improves with an appropriate choice of a target state $\Phi = \sum_{m=0}^{d-1}\lambda_m\ket{mm}$. As mentioned in the main text, we use information from measurements in the standard basis to nominate the target state by identifying the coefficients $\lambda_m$ with:
\be
\lambda_m = \sqrt{\frac{N_{mm}}{\sum_n N_{nn}}},
\ee
where $N_{mn}$ are the coincidence counts obtained when projecting onto states $\ket{m}$ and $\ket{n}$ of the standard LG basis. The values of $\lambda_m$ are then used to construct the set of tilted bases. However, this informed guess of the target state is hindered by the mode-dependent loss introduced by our measurement scheme, which effectively modifies the counts $N_{mn}$ and results in a sub optimal certified fidelity and dimensionality. As an example, we can consider the correlations in the azimuthal component shown in Fig~{\ref{fig:EtaCorrection}}.a (and in Fig.2.b of the main text). The two-photon coincidence counts indicate a state where the probability of detecting pairs of photons in higher order azimuthal modes is larger than the one for lower order modes. This is clearly not the case for the entangled state generated through SPDC, where the entanglement is concentrated around the Gaussian states~\cite{Torres:2003cy}. Such artifact arises from the size of the holograms displayed on the SLMs, optimised for increasing the probability of detecting higher order modes, but in turn, adding loss when projecting onto lower order modes like $LG^{\ell = 0}_{p=0}$. 

To accurately estimate the target state and optimise the performance of the tilted-basis witness, we use the method demonstrated in~\cite{Bavaresco:2018gw} for determining loss factors through the singles detected on each side, and correct the coincidence counts as shown in Fig.~\ref{fig:EtaCorrection}.b. These new $N_{m,n}$ counts are then used for constructing the tilted projectors and certify entanglement.
\begin{figure*}[t!]
    \centering
    \includegraphics[width=0.85\linewidth]{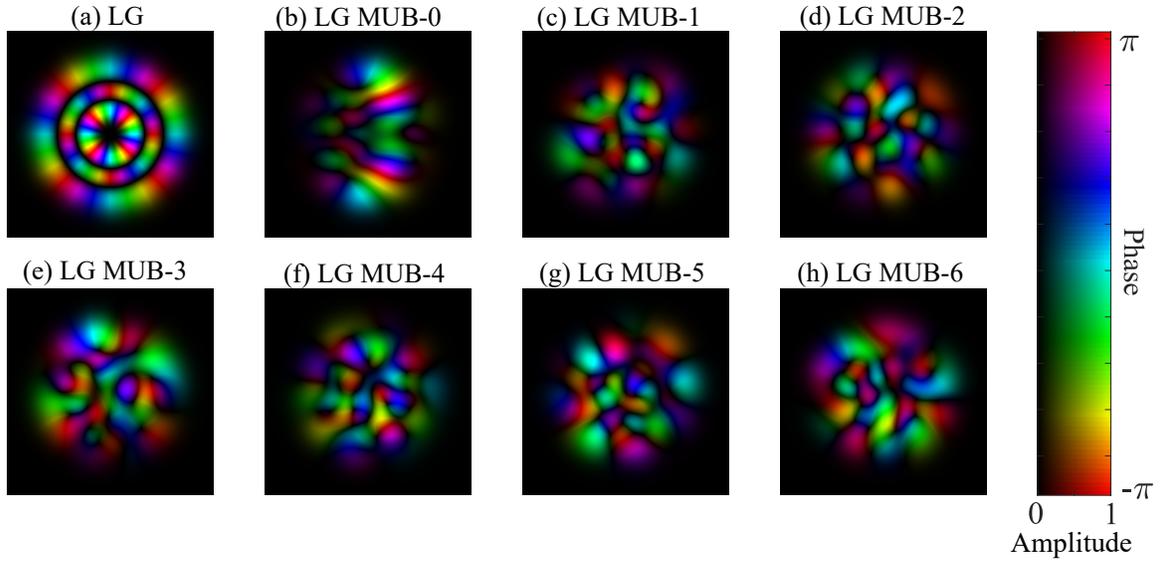}
    \caption{\textbf{Examples of full-field LG and LG MUB modes in dimension $d = 43$}:  Complex amplitudes for the (a) full-field $LG^4_2$ mode and  the (b-h) corresponding $j = d-1^{th}$ LG MUB modes for bases $r = 0$ to $6$.}
    \label{fig:Modes0to7}
\end{figure*}

\begin{figure*}[h!]
    \centering
    \includegraphics[width=0.8 \linewidth]{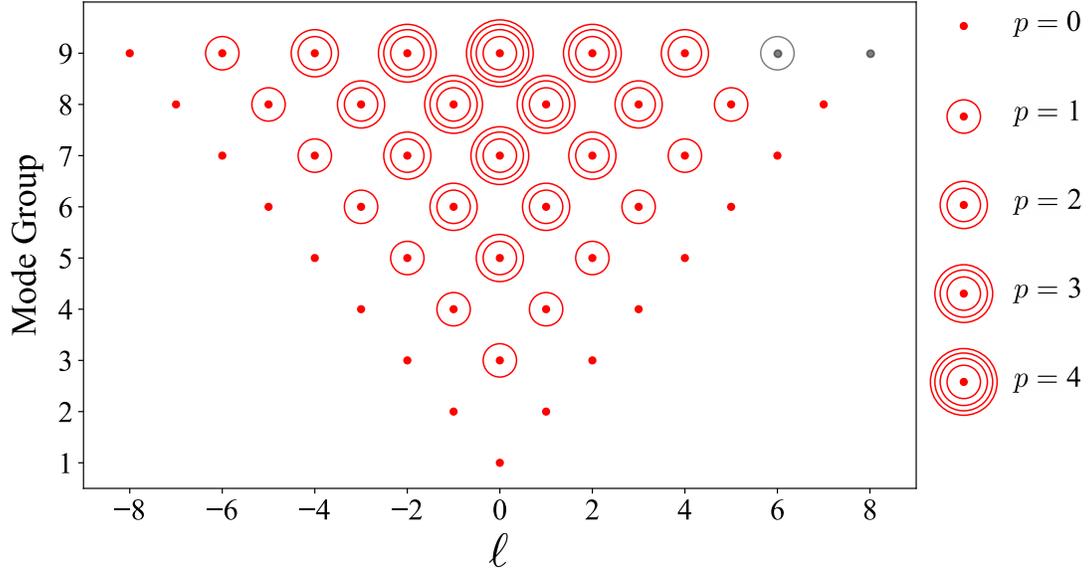}
    \caption{\textbf{Representation of the modes spanning a 43-dimensional space w.~r.~t.~azimuthal mode index $\ell$ for different radial indices $p=0,\dots,4$.}: To perform measurements over a full-field 43-dimensional subspace, we considered modes belonging to 9 different mode groups. Modes in grey are discarded to keep the dimension of the subspace prime.}
    \label{fig:modegroupvsl}
\end{figure*}


 \end{document}